\documentclass[authorversion,screen]{acmart}
 
\AtBeginDocument{%
  }

\setcopyright{rightsretained}

\received{14 March 2025}
\received[accepted]{24 March 2025}

\usepackage{times}                    
\usepackage{enumitem}
\usepackage{wrapfig}

\newcommand{\tool}[0]{\textsc{HybridCollab}}
\newcommand{\arcollab}[0]{\textsc{ARCollab}}

\begin{document}

\title[\tool{}]{\tool{}: 
Unifying In-Person and Remote Collaboration for Cardiovascular Surgical Planning in Mobile Augmented Reality}

\author{Pratham Darrpan Mehta}
\email{pratham@gatech.edu}
\orcid{0000-0001-6650-8489}
\affiliation{%
  \institution{Georgia Institute of Technology}
  \city{Atlanta}
  \state{Georgia}
  \country{USA}
}

\author{Rahul Ozhur Narayanan}
\email{rahulon@gatech.edu}
\orcid{0009-0001-8477-4390}
\affiliation{%
  \institution{Georgia Institute of Technology}
  \city{Atlanta}
  \state{Georgia}
  \country{USA}
}

\author{Vidhi Kulkarni}
\email{vkulkarni65@gatech.edu}
\orcid{0009-0002-3137-6871}
\affiliation{%
  \institution{Georgia Institute of Technology}
  \city{Atlanta}
  \state{Georgia}
  \country{USA}
}

\author{Timothy C. Slesnick}
\orcid{0009-0000-6532-9788}
\affiliation{%
  \institution{Children's Healthcare of Atlanta}
  \city{Atlanta}
  \country{USA}}
\email{SlesnickT@kidsheart.com}

\author{Fawwaz Shaw}
\orcid{0000-0002-9873-8390}
\affiliation{%
  \institution{Children's Healthcare of Atlanta}
  \city{Atlanta}
  \country{USA}}
\email{Fawwaz.Shaw@choa.org}

\author{Duen Horng Chau}
\orcid{0000-0001-9824-3323}
\affiliation{%
  \institution{Georgia Institute of Technology}
  \city{Atlanta}
  \country{USA}}
\email{polo@gatech.edu}

\renewcommand{\shortauthors}{Mehta et al.}

\begin{abstract}
 Surgical planning for congenital heart disease traditionally relies on collaborative group examinations of a patient's 3D-printed heart model, a process that lacks flexibility and accessibility.
 While mobile augmented reality (AR) offers a promising alternative with its portability and familiar interaction gestures, existing solutions limit collaboration to users in the same physical space. 
 We developed \tool{}, the first iOS AR application that introduces a novel paradigm that enables both in-person and remote medical teams to interact with a shared AR heart model in a single surgical planning session. 
 For example, a team of two doctors in one hospital room can collaborate in real time with another team in a different hospital.
 Our approach is the first to leverage Apple's GameKit service for surgical planning, ensuring an identical collaborative experience for all participants, regardless of location. 
 Additionally, co-located users can interact with the same anchored heart model in their shared physical space. 
 By bridging the gap between remote and in-person collaboration across medical teams, \tool{} has the potential for significant real-world impact, streamlining communication and enhancing the effectiveness of surgical planning. Watch the demo: \url{https://youtu.be/hElqJYDuvLM}.
 
\end{abstract}



\begin{teaserfigure}
  \includegraphics[width=\textwidth]{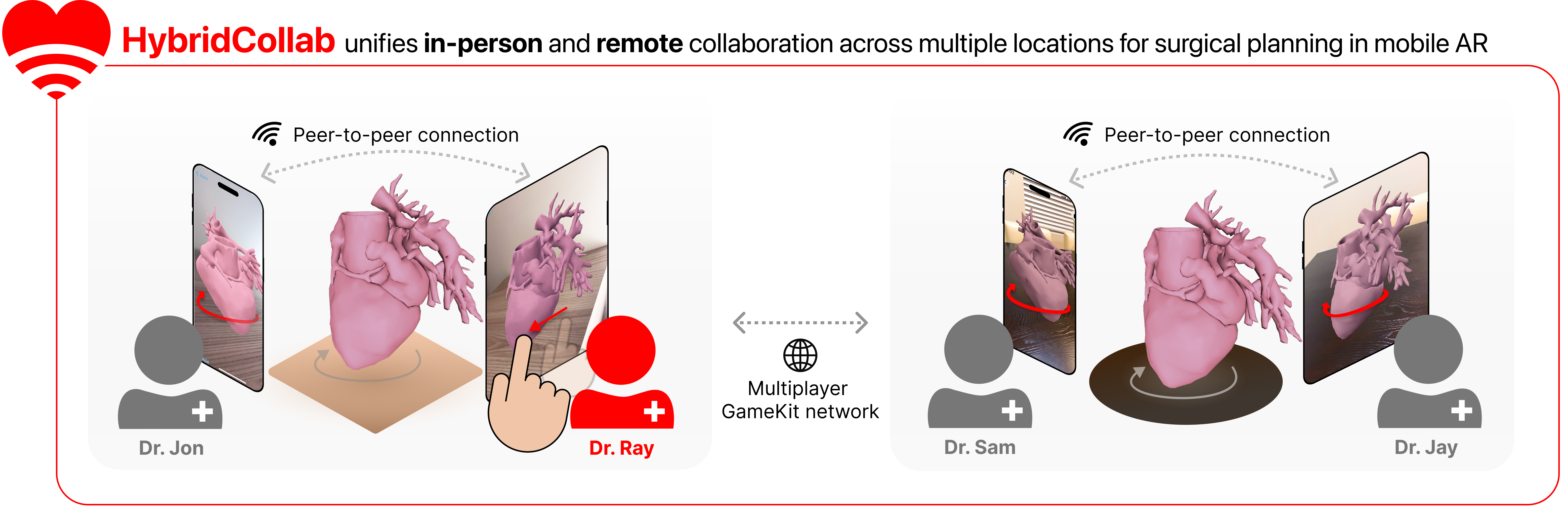}
  \caption{\tool{} unifies in-person and remote collaboration for cardiovascular surgical planning in mobile augmented reality (AR). 
  Dr. Ray and Dr. Jon are in one hospital room, while Dr. Sam and Dr. Jay are in another, all collaborating in a single shared AR session.
  When Dr. Ray interacts with the heart model (as shown by the finger gesture), the heart's state is updated across all connected devices, as indicated by the red arrows on all devices. 
  }
  \Description{Teaser image of HybridCollab}
  \label{fig:teaser}
\end{teaserfigure}

\maketitle

\section{Introduction}
\begin{wrapfigure}{r}{0.2\textwidth}
\includegraphics[width=\linewidth]{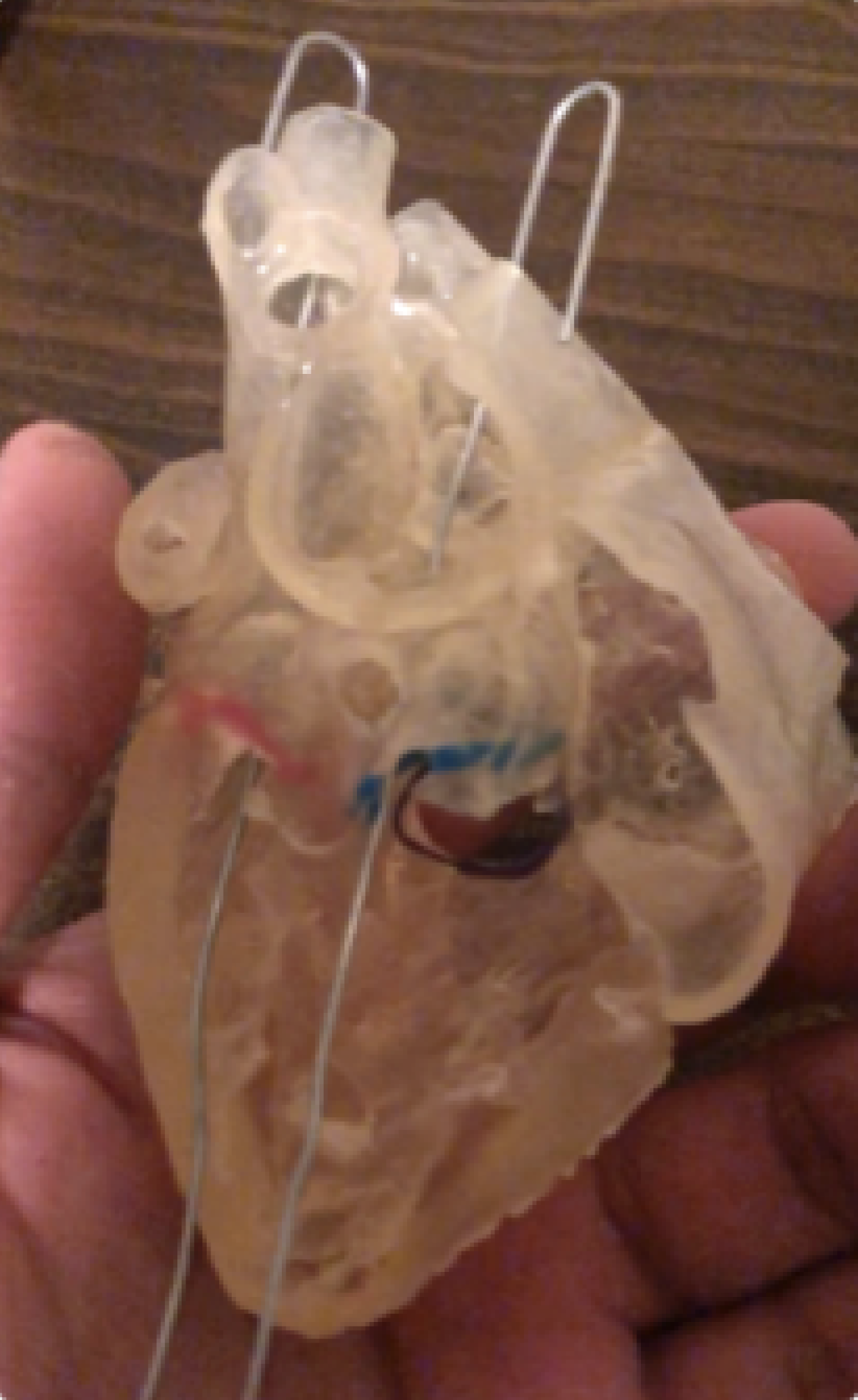} 
\caption{Example of a 3D-printed heart model.}
\label{fig:heart-cutout}
\end{wrapfigure}

Surgical planning for congenital heart disease (CHD) treatment typically involves collaborative examination of physical three-dimensional (3D) heart models, 
as shown in Fig. \ref{fig:heart-cutout}, 
created through medical imaging
\cite{riggs_3d-printed_2018}. 
However, producing these physical models requires hours of processing and specialized, often inaccessible, 3D printers \cite{yoo_3d_2021}. 
Extended reality (XR) technologies, including virtual reality (VR), mixed reality (MR), and augmented reality (AR), offer promising alternatives for creating more adaptable and collaborative planning environments \cite{Sun2019PersonalizedTP, Gr2018AdaptiveA, Kappanayil2017ThreedimensionalprintedCP, yoo_3d_2021, sun_3d_2022, schott_vrar_2021, zhang_directx-based_2022, dass_augmenting_2018, hirzle_critical_2021}.

Our prior work, CardiacAR, demonstrated the potential of mobile AR in surgical planning due to its portability, familiar gestures, and ease of use \cite{Leo21, yang_evaluating_2022}. 
While CardiacAR enabled interactive visualization,
it was missing a key part of the planning process: collaboration. 
\arcollab{} was then developed as the first multi-user mobile AR surgical planning tool \cite{mehta2024arcollab}. 
It allowed surgeons and cardiologists to collaboratively interact with a 3D heart model in the same physical space. 

Through usability studies, we found that while \arcollab{} enabled in-person collaboration for surgical planning, it required all users to be in the same physical space,
a constraint that poses a barrier to adoption \cite{mehta_multi-user_2024}. 
Surgeons and cardiologists emphasized that overcoming this limitation is essential as medical teams are often in different physical locations \cite{mehta_multi-user_2024}.

To address the above research gaps, our ongoing work makes the following contributions:

\begin{enumerate}[topsep=2pt, itemsep=0mm, parsep=3pt, leftmargin=10pt]

    \item \textbf{\tool{}, the first approach that 
    unifies in-person and remote
    collaboration in mobile AR for iOS.
    } 
    Developed in collaboration with \textit{Children's Healthcare of Atlanta} (CHOA), \tool{} breaks new ground as the first system to seamlessly integrate both physically co-located and geographically distributed medical teams for surgical planning.
    We pioneer this novel paradigm by uniquely adapting and combining multiple state-of-the-art frameworks: repurposing a real-time multiplayer network to support up to 16 medical professionals simultaneously, customizing a peer-to-peer networking service for spatial coordination among co-located devices, and leveraging advanced AR and graphics frameworks for interactive visualization of complex cardiac anatomy across distributed environments.
    Consider a scenario where Doctors Ray and Jon are collaborating in one hospital while Doctors Sam and Jay are in another facility (Fig. \ref{fig:teaser}). 
    When all four join a shared \tool{} session, each co-located team anchors a 3D heart model in their physical space. 
    As Dr. Ray manipulates the model, these changes instantly propagate across the network to all connected devices, ensuring a synchronized visualization for the entire team, regardless of location. Watch a demo: \url{https://youtu.be/hElqJYDuvLM}.

    \item 
    \textbf{Expanding access to surgical planning 
    through mobile platforms}. 
    \tool{} is developed for widely available iOS devices, leveraging their portability and ease of use to facilitate convenient and efficient collaboration \cite{dass_augmenting_2018}. 
    Using Apple's TestFlight service, we can rapidly deploy updated versions of \tool{} for usability testing and long-term evaluations.
\end{enumerate}

\section{Design Goals}
\label{sec:goals}
Over the past three years, we have worked closely with cardiothoracic surgeons and cardiologists from CHOA to develop \tool{}, refining it through ongoing consultation. In this process, we have identified three primary design goals:
\begin{enumerate}[topsep=1pt, itemsep=2mm, parsep=1pt, leftmargin=19pt, label=\textnormal{\textbf{G\arabic*.}}, ref=G\arabic*]

    \item \label{goal:hybrid-collab} 
    \textbf{Enable interactive hybrid collaboration regardless of location.}
    Our prior work demosntrated that multi-user mobile AR significantly enhances collaboration for doctors in the same physical space during surgical planning \cite{mehta_multi-user_2024}. 
    However, no existing tool supports hybrid surgical planning for doctors in different locations. 
    We aim to create a tool that enables seamless collaboration between in-person and remote medical teams, ensuring an equally collaborative experience for all participants, regardless of their physical setting.

    \item \label{goal:low-latency} 
    \textbf{Develop low-latency technology for collaborative sessions and efficient communication}
    Real-time interaction is essential for effective collaboration \cite{He02102017}.
    To achieve low-latency and ensure rapid deployment, we aim to adapt native iOS frameworks for direct device-to-device communication, reducing the need for server maintenance.

    \item \label{goal:effective-comms}
    \textbf{Support familiar interaction and communication mechanisms across all settings.}
    Easy-to-use gestures and familiar interactions like panning, tapping, and pinching are crucial for system adoption \cite{DAKULAGI202563}.
    However, when collaborators are in different physical locations, traditional pointing and verbal descriptions may fail to convey spatial information about anatomical structures effectively.
    To address this, we aim to implement intuitive finger gestures and spatial referencing tools to bridge communication gaps across both in-person and remote settings.
\end{enumerate}

\section{System Design and Implementation}

\label{sec:remote-collab}
\subsection{Creating a Collaborative Session via GameKit (\ref{goal:hybrid-collab}, \ref{goal:low-latency})}
\begin{wrapfigure}{r}{0.25\textwidth}
\centering
\includegraphics[width=\linewidth]{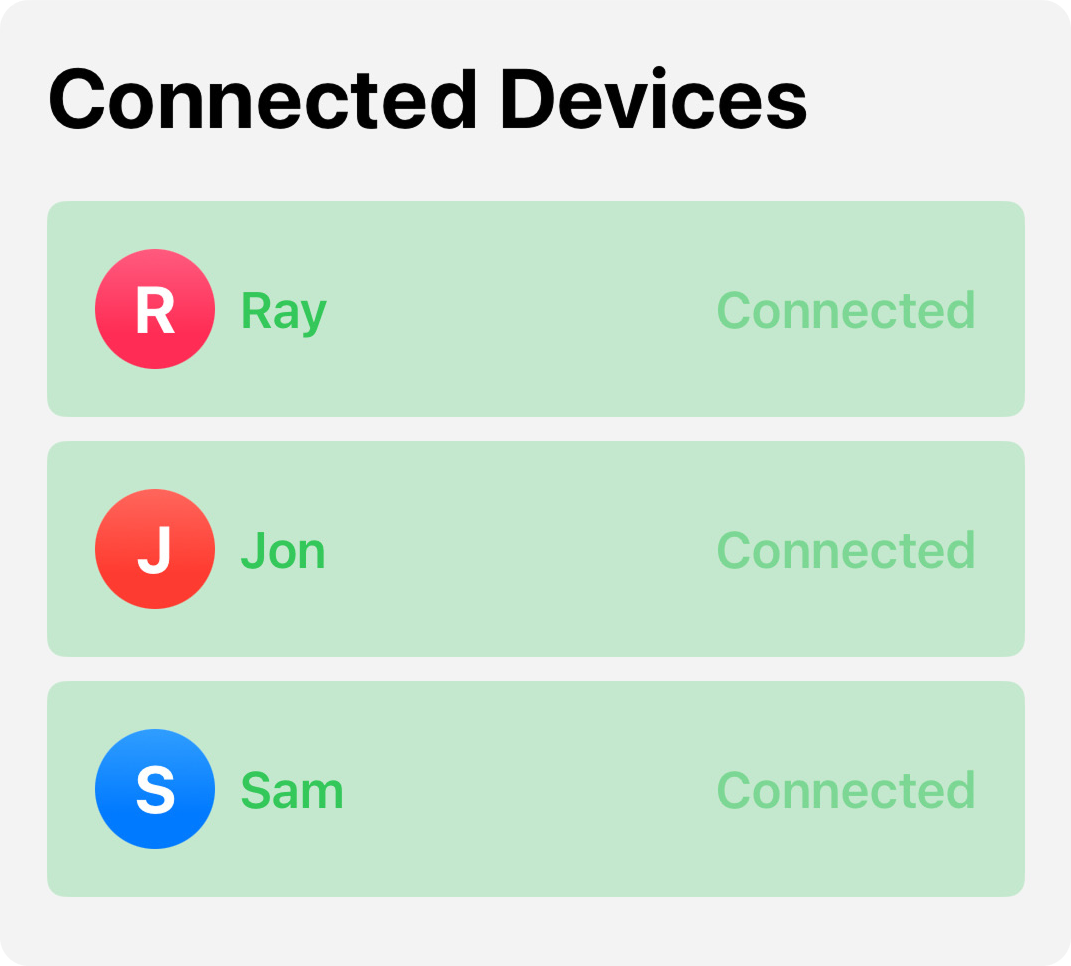} 
\caption{\tool{} users connected in a session.}
\label{fig:connected-devices}
\end{wrapfigure}

\tool{} leverages GameKit, Apple's multiplayer service, to connect users across various physical locations. 
Our implementation transforms this gaming-oriented framework into a  medical collaboration platform that supports up to 16 simultaneous users, enabling real-time, low-latency communication crucial for precise surgical planning.
Our approach leverages Apple's robust network infrastructure, eliminating the need for custom server deployment and maintenance while ensuring consistent performance.

\subsubsection{For users at different physical locations}

When a user opens \tool{} to join a session remotely, the network begins searching for other available devices.
Users are automatically matched and displayed on all devices, as shown in Fig. \ref{fig:connected-devices}.
Once the session starts, an AR view is presented to all connected users, each of whom scans their own physical environment,
allowing ARKit to gather relevant data about horizontal surfaces in the surroundings. 
Each user can then tap the screen to place the 3D heart model on a surface of their choice.

After all users have placed the heart model, they can begin interacting with it using \tool{}'s built-in gestures. 
Building on our previous implementations, we have optimized the transformation encoding process to minimize data transmission across GameKit's network, enabling near-instantaneous synchronization of the heart model's state across all devices regardless of physical distance.

\tool{} also integrates Apple SharePlay support, allowing up to 16 users to share, view, and manipulate the heart model together while connecting via a FaceTime call \cite{inc_shareplay_nodate}. This enables remote users to communicate effectively without the need for external tools.

\subsubsection{For users sharing the same physical space}
When multiple users are in the same physical space, such as a room, 
the app automatically broadcasts and connects to nearby devices using Multipeer Connectivity, Apple's peer-to-peer networking service.
As in our previous work, the users can scan their shared physical environment, allowing their devices to detect common feature points and synchronize their surroundings \cite{mehta2024arcollab}. 
This ensures that each group of in-person users see a single heart model anchored to a common point in their environment.

To maintain consistent collaboration across both remote and in-person users, all transformations are encoded and shared via GameKit's multiplayer network. This is necessary because peer-to-peer networks are localized to groups of users within the same physical space.

\begin{wrapfigure}{r}{0.25\textwidth}
\centering
\includegraphics[width=\linewidth]{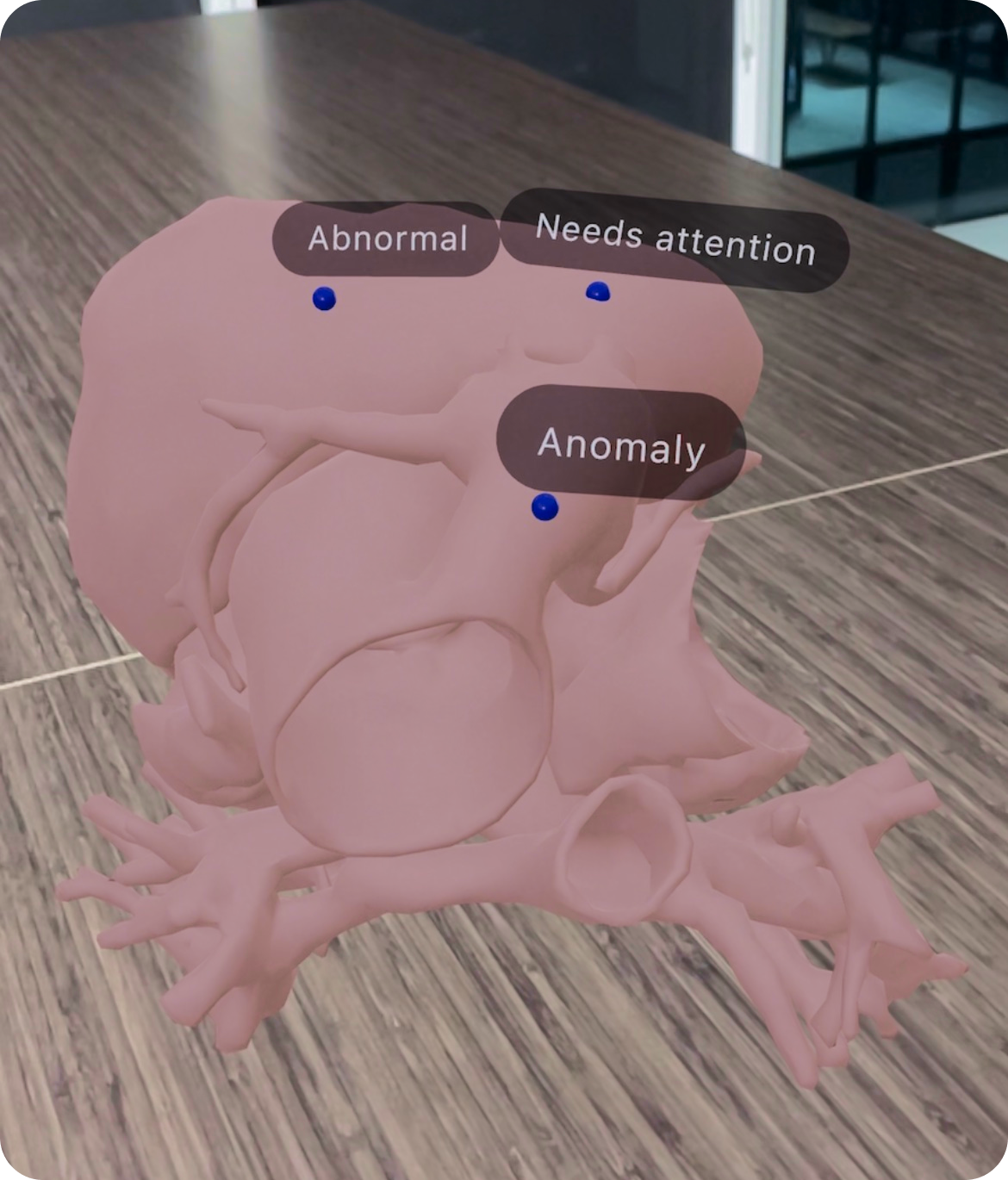} 
\caption{Example annotations.}
\label{fig:annotation-screenshot}
\end{wrapfigure}

\subsection{Virtual Annotation and Slicing (\ref{goal:hybrid-collab}, \ref{goal:effective-comms})}

Through prior usability studies, we have found that labeling specific regions of the heart model enhances communication \cite{yang_evaluating_2022}.
As a result, we incorporate this feature into \tool{}.
When a user taps on the screen, the application performs raycasting to identify where an annotation should be placed \cite{Leo21}.
Each annotation is represented by a small sphere and accompanying text,
as shown in Fig. \ref{fig:annotation-screenshot}.
All annotations
are shared and updated across all connected devices in the session.

We are also extending this feature to include a virtual pointing mechanism to highlight specific regions of the heart model. 
When a user taps the screen, a uniquely colored ray extends from their device through the AR environment and intersects with the heart model, creating a visual indicator visible to all participants.
Unlike annotations, which are persistent, this pointing mechanism is temporary and designed for dynamic discussions.

Finally, \tool{} builds on our prior work to support multi-user omnidirectional slicing, allowing users to perform reversible, cross-sectional slices across any plane to  view the inner morphology of the heart model. 
Traditionally, this is done with physical models, but it often results in irreversible alterations, limiting doctors's ability to explore different cross-sections \cite{riggs_3d-printed_2018,Kappanayil2017ThreedimensionalprintedCP,yoo_3d_2021}.

\section{Conclusion and Future Work}
We have presented \tool{}, our ongoing research that enables a new paradigm in collaborative surgical planning by unifying in-person and remote collaboration through mobile AR. 
By facilitating hybrid collaboration regardless of physical location, we address the critical need for flexible participation in surgical planning while maintaining the interactive benefits of mobile AR technology.
Our implementation integrates multiple state-of-the-art frameworks to create a seamless experience for medical teams across different physical locations. 
Moving forward, we plan to conduct comprehensive usability studies with surgeons and cardiologists from CHOA to assess \tool{}'s effectiveness in real-world settings. 
Through iterative testing and refinement of communication features, we aim to further bridge the gap between remote and in-person participants, ultimately enhancing collaborative surgical planning outcomes.

\newpage

\bibliographystyle{ACM-Reference-Format}
\bibliography{bibliography}


\begin{thebibliography}{17}


\ifx \showCODEN    \undefined \def \showCODEN     #1{\unskip}     \fi
\ifx \showISBNx    \undefined \def \showISBNx     #1{\unskip}     \fi
\ifx \showISBNxiii \undefined \def \showISBNxiii  #1{\unskip}     \fi
\ifx \showISSN     \undefined \def \showISSN      #1{\unskip}     \fi
\ifx \showLCCN     \undefined \def \showLCCN      #1{\unskip}     \fi
\ifx \shownote     \undefined \def \shownote      #1{#1}          \fi
\ifx \showarticletitle \undefined \def \showarticletitle #1{#1}   \fi
\ifx \showURL      \undefined \def \showURL       {\relax}        \fi
\providecommand\bibfield[2]{#2}
\providecommand\bibinfo[2]{#2}
\providecommand\natexlab[1]{#1}
\providecommand\showeprint[2][]{arXiv:#2}

\bibitem[Dakulagi et~al\mbox{.}(2025)]%
        {DAKULAGI202563}
\bibfield{author}{\bibinfo{person}{Veerendra Dakulagi}, \bibinfo{person}{Kim~Ho Yeap}, \bibinfo{person}{Humaira Nisar}, \bibinfo{person}{Rohini Dakulagi}, \bibinfo{person}{G~N Basavaraj}, {and} \bibinfo{person}{Miguel~Villagómez Galindo}.} \bibinfo{year}{2025}\natexlab{}.
\newblock \showarticletitle{Chapter 4 - An overview of techniques and best practices to create intuitive and user-friendly human-machine interfaces}.
\newblock In \bibinfo{booktitle}{\emph{Artificial Intelligence and Multimodal Signal Processing in Human-Machine Interaction}}, \bibfield{editor}{\bibinfo{person}{Abdulhamit Subasi}, \bibinfo{person}{Saeed~Mian Qaisar}, {and} \bibinfo{person}{Humaira Nisar}} (Eds.). \bibinfo{publisher}{Academic Press}, \bibinfo{pages}{63--77}.
\newblock
\showISBNx{978-0-443-29150-0}
\href{https://doi.org/10.1016/B978-0-443-29150-0.00002-0}{doi:\nolinkurl{10.1016/B978-0-443-29150-0.00002-0}}


\bibitem[Dass et~al\mbox{.}(2018)]%
        {dass_augmenting_2018}
\bibfield{author}{\bibinfo{person}{Nathan Dass}, \bibinfo{person}{Joonyoung Kim}, \bibinfo{person}{Sam Ford}, \bibinfo{person}{Sudeep Agarwal}, {and} \bibinfo{person}{Duen Horng~(Polo) Chau}.} \bibinfo{year}{2018}\natexlab{}.
\newblock \showarticletitle{Augmenting {Coding}: {Augmented} {Reality} for {Learning} {Programming}}. In \bibinfo{booktitle}{\emph{Proceedings of the {Sixth} {International} {Symposium} of {Chinese} {CHI}}}. \bibinfo{publisher}{ACM}, \bibinfo{address}{Montreal QC Canada}, \bibinfo{pages}{156--159}.
\newblock
\showISBNx{978-1-4503-6508-6}
\href{https://doi.org/10.1145/3202667.3202695}{doi:\nolinkurl{10.1145/3202667.3202695}}


\bibitem[Gr et~al\mbox{.}(2018)]%
        {Gr2018AdaptiveA}
\bibfield{author}{\bibinfo{person}{Padmaja Gr}, \bibinfo{person}{He}, \bibinfo{person}{Sreenivasa~Reddy Edara}, {and} \bibinfo{person}{Vasumathi Devara}.} \bibinfo{year}{2018}\natexlab{}.
\newblock \showarticletitle{Adaptive analysis \& reconstruction of 3D DICOM images using enhancement based SBIR algorithm over MRI}.
\newblock \bibinfo{journal}{\emph{Biomedical Research-tokyo}}  \bibinfo{volume}{29} (\bibinfo{year}{2018}), \bibinfo{pages}{644--653}.
\newblock


\bibitem[He and Huang(2017)]%
        {He02102017}
\bibfield{author}{\bibinfo{person}{Jinxia He} {and} \bibinfo{person}{Xiaoxia Huang}.} \bibinfo{year}{2017}\natexlab{}.
\newblock \showarticletitle{Collaborative Online Teamwork: Exploring Students' Satisfaction and Attitudes with Google Hangouts as a Supplementary Communication Tool}.
\newblock \bibinfo{journal}{\emph{Journal of Research on Technology in Education}} \bibinfo{volume}{49}, \bibinfo{number}{3-4} (\bibinfo{year}{2017}), \bibinfo{pages}{149--160}.
\newblock
\href{https://doi.org/10.1080/15391523.2017.1327334}{doi:\nolinkurl{10.1080/15391523.2017.1327334}}
\showeprint{https://doi.org/10.1080/15391523.2017.1327334}


\bibitem[Hirzle et~al\mbox{.}(2021)]%
        {hirzle_critical_2021}
\bibfield{author}{\bibinfo{person}{Teresa Hirzle}, \bibinfo{person}{Maurice Cordts}, \bibinfo{person}{Enrico Rukzio}, \bibinfo{person}{Jan Gugenheimer}, {and} \bibinfo{person}{Andreas Bulling}.} \bibinfo{year}{2021}\natexlab{}.
\newblock \showarticletitle{A {Critical} {Assessment} of the {Use} of {SSQ} as a {Measure} of {General} {Discomfort} in {VR} {Head}-{Mounted} {Displays}}. In \bibinfo{booktitle}{\emph{Proceedings of the 2021 {CHI} {Conference} on {Human} {Factors} in {Computing} {Systems}}}. \bibinfo{publisher}{ACM}, \bibinfo{address}{Yokohama Japan}, \bibinfo{pages}{1--14}.
\newblock
\showISBNx{978-1-4503-8096-6}
\href{https://doi.org/10.1145/3411764.3445361}{doi:\nolinkurl{10.1145/3411764.3445361}}


\bibitem[Inc({[n.\,d.]})]%
        {inc_shareplay_nodate}
\bibfield{author}{\bibinfo{person}{Apple Inc}.} \bibinfo{year}{[n.\,d.]}\natexlab{}.
\newblock \bibinfo{title}{{SharePlay}}.
\newblock
\urldef\tempurl%
\url{https://developer.apple.com/shareplay/}
\showURL{%
\tempurl}


\bibitem[Kappanayil et~al\mbox{.}(2017)]%
        {Kappanayil2017ThreedimensionalprintedCP}
\bibfield{author}{\bibinfo{person}{Mahesh Kappanayil}, \bibinfo{person}{Nageshwara~Rao Koneti}, \bibinfo{person}{Rajesh~R Kannan}, \bibinfo{person}{Brijesh~Parayaru Kottayil}, {and} \bibinfo{person}{Krishna Kumar}.} \bibinfo{year}{2017}\natexlab{}.
\newblock \showarticletitle{Three-dimensional-printed cardiac prototypes aid surgical decision-making and preoperative planning in selected cases of complex congenital heart diseases: Early experience and proof of concept in a resource-limited environment}.
\newblock \bibinfo{journal}{\emph{Annals of Pediatric Cardiology}}  \bibinfo{volume}{10} (\bibinfo{year}{2017}), \bibinfo{pages}{117 -- 125}.
\newblock


\bibitem[Leo et~al\mbox{.}(2021)]%
        {Leo21}
\bibfield{author}{\bibinfo{person}{Jonathan Leo}, \bibinfo{person}{Zhiyan Zhou}, \bibinfo{person}{H. Yang}, \bibinfo{person}{Megan Dass}, \bibinfo{person}{Anish Upadhayay}, \bibinfo{person}{Timothy~C. Slesnick}, \bibinfo{person}{Fawwaz Shaw}, {and} \bibinfo{person}{Duen~Horng Chau}.} \bibinfo{year}{2021}\natexlab{}.
\newblock \showarticletitle{Interactive Cardiovascular Surgical Planning via Augmented Reality}.
\newblock \bibinfo{journal}{\emph{Asian CHI Symposium 2021}} (\bibinfo{year}{2021}).
\newblock


\bibitem[Mehta et~al\mbox{.}(2024a)]%
        {mehta2024arcollab}
\bibfield{author}{\bibinfo{person}{Pratham Mehta}, \bibinfo{person}{Harsha Karanth}, \bibinfo{person}{Haoyang Yang}, \bibinfo{person}{Timothy Slesnick}, \bibinfo{person}{Fawwaz Shaw}, {and} \bibinfo{person}{Duen~Horng Chau}.} \bibinfo{year}{2024}\natexlab{a}.
\newblock \bibinfo{title}{ARCollab: Towards Multi-User Interactive Cardiovascular Surgical Planning in Mobile Augmented Reality}.
\newblock
\showeprint[arxiv]{2402.05075}~[cs.HC]


\bibitem[Mehta et~al\mbox{.}(2024b)]%
        {mehta_multi-user_2024}
\bibfield{author}{\bibinfo{person}{Pratham Mehta}, \bibinfo{person}{Rahul Narayanan}, \bibinfo{person}{Harsha Karanth}, \bibinfo{person}{Haoyang Yang}, \bibinfo{person}{Timothy~C. Slesnick}, \bibinfo{person}{Fawwaz Shaw}, {and} \bibinfo{person}{Duen~Horng Chau}.} \bibinfo{year}{2024}\natexlab{b}.
\newblock \showarticletitle{Multi-{User} {Mobile} {Augmented} {Reality} for {Cardiovascular} {Surgical} {Planning}}. In \bibinfo{booktitle}{\emph{2024 {IEEE} {Visualization} and {Visual} {Analytics} ({VIS})}}. \bibinfo{pages}{201--205}.
\newblock
\href{https://doi.org/10.1109/VIS55277.2024.00048}{doi:\nolinkurl{10.1109/VIS55277.2024.00048}}
\newblock
\shownote{ISSN: 2771-9553}.


\bibitem[Riggs et~al\mbox{.}(2018)]%
        {riggs_3d-printed_2018}
\bibfield{author}{\bibinfo{person}{Kyle~W. Riggs}, \bibinfo{person}{Gavin Dsouza}, \bibinfo{person}{John~T. Broderick}, \bibinfo{person}{Ryan~A. Moore}, {and} \bibinfo{person}{David L.~S. Morales}.} \bibinfo{year}{2018}\natexlab{}.
\newblock \showarticletitle{{3D}-printed models optimize preoperative planning for pediatric cardiac tumor debulking}.
\newblock \bibinfo{journal}{\emph{Translational Pediatrics}} \bibinfo{volume}{7}, \bibinfo{number}{3} (\bibinfo{date}{July} \bibinfo{year}{2018}), \bibinfo{pages}{196--202}.
\newblock
\showISSN{2224-4344}
\href{https://doi.org/10.21037/tp.2018.06.01}{doi:\nolinkurl{10.21037/tp.2018.06.01}}


\bibitem[Schott et~al\mbox{.}(2021)]%
        {schott_vrar_2021}
\bibfield{author}{\bibinfo{person}{Danny Schott}, \bibinfo{person}{Patrick Saalfeld}, \bibinfo{person}{Gerd Schmidt}, \bibinfo{person}{Fabian Joeres}, \bibinfo{person}{Christian Boedecker}, \bibinfo{person}{Florentine Huettl}, \bibinfo{person}{Hauke Lang}, \bibinfo{person}{Tobias Huber}, \bibinfo{person}{Bernhard Preim}, {and} \bibinfo{person}{Christian Hansen}.} \bibinfo{year}{2021}\natexlab{}.
\newblock \showarticletitle{A {VR}/{AR} {Environment} for {Multi}-{User} {Liver} {Anatomy} {Education}}. In \bibinfo{booktitle}{\emph{2021 {IEEE} {Virtual} {Reality} and {3D} {User} {Interfaces} ({VR})}}. \bibinfo{pages}{296--305}.
\newblock
\href{https://doi.org/10.1109/VR50410.2021.00052}{doi:\nolinkurl{10.1109/VR50410.2021.00052}}
\newblock
\shownote{ISSN: 2642-5254}.


\bibitem[Sun et~al\mbox{.}(2019)]%
        {Sun2019PersonalizedTP}
\bibfield{author}{\bibinfo{person}{Zhonghua Sun}, \bibinfo{person}{Ivan Wen~Wen Lau}, \bibinfo{person}{Yin~How Wong}, {and} \bibinfo{person}{Chai~Hong Yeong}.} \bibinfo{year}{2019}\natexlab{}.
\newblock \showarticletitle{Personalized Three-Dimensional Printed Models in Congenital Heart Disease}.
\newblock \bibinfo{journal}{\emph{Journal of Clinical Medicine}}  \bibinfo{volume}{8} (\bibinfo{year}{2019}).
\newblock


\bibitem[Sun and Wee(2022)]%
        {sun_3d_2022}
\bibfield{author}{\bibinfo{person}{Zhonghua Sun} {and} \bibinfo{person}{Cleo Wee}.} \bibinfo{year}{2022}\natexlab{}.
\newblock \showarticletitle{{3D} {Printed} {Models} in {Cardiovascular} {Disease}: {An} {Exciting} {Future} to {Deliver} {Personalized} {Medicine}}.
\newblock \bibinfo{journal}{\emph{Micromachines}} \bibinfo{volume}{13}, \bibinfo{number}{10} (\bibinfo{date}{Oct.} \bibinfo{year}{2022}), \bibinfo{pages}{1575}.
\newblock
\showISSN{2072-666X}
\href{https://doi.org/10.3390/mi13101575}{doi:\nolinkurl{10.3390/mi13101575}}
\newblock
\shownote{Number: 10 Publisher: Multidisciplinary Digital Publishing Institute}.


\bibitem[Yang et~al\mbox{.}(2022)]%
        {yang_evaluating_2022}
\bibfield{author}{\bibinfo{person}{Haoyang Yang}, \bibinfo{person}{Pratham~Darrpan Mehta}, \bibinfo{person}{Jonathan Leo}, \bibinfo{person}{Zhiyan Zhou}, \bibinfo{person}{Megan Dass}, \bibinfo{person}{Anish Upadhayay}, \bibinfo{person}{Timothy~C. Slesnick}, \bibinfo{person}{Fawwaz Shaw}, \bibinfo{person}{Amanda Randles}, {and} \bibinfo{person}{Duen~Horng Chau}.} \bibinfo{year}{2022}\natexlab{}.
\newblock \bibinfo{title}{Evaluating {Cardiovascular} {Surgical} {Planning} in {Mobile} {Augmented} {Reality}}.
\newblock
\href{https://doi.org/10.48550/arXiv.2208.10639}{doi:\nolinkurl{10.48550/arXiv.2208.10639}}
\newblock
\shownote{arXiv:2208.10639 [cs]}.


\bibitem[Yoo et~al\mbox{.}(2021)]%
        {yoo_3d_2021}
\bibfield{author}{\bibinfo{person}{Shi~Joon Yoo}, \bibinfo{person}{Nabil Hussein}, \bibinfo{person}{Brandon Peel}, \bibinfo{person}{John Coles}, \bibinfo{person}{Glen S.~van Arsdell}, \bibinfo{person}{Osami Honjo}, \bibinfo{person}{Christoph Haller}, \bibinfo{person}{Christopher~Z. Lam}, \bibinfo{person}{Mike Seed}, {and} \bibinfo{person}{David Barron}.} \bibinfo{year}{2021}\natexlab{}.
\newblock \showarticletitle{{3D} {Modeling} and {Printing} in {Congenital} {Heart} {Surgery}: {Entering} the {Stage} of {Maturation}}.
\newblock \bibinfo{journal}{\emph{Frontiers in Pediatrics}}  \bibinfo{volume}{9} (\bibinfo{year}{2021}).
\newblock
\showISSN{2296-2360}
\urldef\tempurl%
\url{https://www.frontiersin.org/article/10.3389/fped.2021.621672}
\showURL{%
\tempurl}


\bibitem[Zhang et~al\mbox{.}(2022)]%
        {zhang_directx-based_2022}
\bibfield{author}{\bibinfo{person}{Menghe Zhang}, \bibinfo{person}{Weichen Liu}, \bibinfo{person}{Nadir Weibel}, {and} \bibinfo{person}{Jurgen Schulze}.} \bibinfo{year}{2022}\natexlab{}.
\newblock \bibinfo{title}{A {DirectX}-{Based} {DICOM} {Viewer} for {Multi}-{User} {Surgical} {Planning} in {Augmented} {Reality}}.
\newblock
\urldef\tempurl%
\url{http://arxiv.org/abs/2210.14349}
\showURL{%
\tempurl}
\newblock
\shownote{arXiv:2210.14349 [cs]}.


\end{thebibliography}

\end{document}